\documentclass{aa}
\usepackage{graphicx,natbib}
\usepackage{txfonts}
\bibpunct{(}{)}{;}{a}{}{,}

\begin{document}

\title{Effective area calibration of the reflection
grating spectrometers of XMM-Newton. II. X-ray spectroscopy of DA white dwarfs}

\author{Jelle Kaastra\inst{1,2}
\and T. Lanz\inst{3}
\and I. Hubeny\inst{4}
\and F.B.S. Paerels\inst{5}
}

\offprints{J. Kaastra}
\date{\today}

\institute{SRON Netherlands Institute for Space Research, Sorbonnelaan 2,
           3584 CA Utrecht, the Netherlands 
	   \and
	   Sterrenkundig Instituut, Universiteit Utrecht, 
P.O. Box 80000, 3508 TA Utrecht, The Netherlands
	\and
	Department of Astronomy, University of Maryland,  
	College Park, MD 20742, USA
	\and 
	Department of Astronomy and Steward Observatory, The University
of Arizona, Tucson, AZ 85721
	\and
	Department of Astronomy and Columbia Astrophysics Laboratory,
Columbia University, 550 West 120th Street, New York, NY 10027, USA
	 }

\newcommand{\Msun}{M$_{\odot}$}
\hyphenation{Ed-ding-ton}

\abstract
% context heading (optional)
% {} leave it empty if necessary  
{ White dwarf spectra have been widely used as a calibration source 
for X-ray and EUV
instruments. The in-flight effective area calibration of the reflection
grating spectrometers (RGS) of XMM-Newton depend upon the availability
of reliable calibration sources. }
% aims heading (mandatory)
{ We investigate how well these white dwarf spectra can be used as standard
candles at the lowest X-ray energies in order to gauge the absolute effective
area scale of X-ray instruments.} 
% methods heading (mandatory) 
{ We calculate a grid of model atmospheres for Sirius~B and HZ~43A, and adjust
the parameters using several constraints until the ratio of the spectra of
both stars agrees with the ratio as observed by the low energy
transmission grating spectrometer (LETGS) of Chandra. This ratio is independent
of any errors in the effective area of the LETGS. }
% results heading (mandatory) 
{ We find that we can constrain the absolute X-ray spectrum of both stars with
better than 5~\% accuracy. The best-fit model for both stars is close to a
pure hydrogen atmosphere, and we put tight limits to the amount of helium
or the thickness of a hydrogen layer in both stars. Our upper limit to the
helium abundance in Sirius~B is 4 times below the previous detection based on
EUVE data. We also find that our results are sensitive to the adopted
cut-off in the Lyman pseudo-continuum opacity in Sirius~B. We get best agreement
with a long wavelength cut-off.} 
% conclusions heading (optional), leave it empty if necessary  
{ White dwarf model atmospheres can be used to derive the effective area
of X-ray spectrometers in the lowest energy band. An accuracy of 3--4~\%
in the absolute effective area can be achieved.}


\titlerunning{X-ray spectroscopy of DA white dwarfs}
\authorrunning{J.S. Kaastra et al.}

\maketitle

\section{Introduction}

DA white dwarfs such as \object{Sirius~B} and \object{HZ~43A} can be used very
well as calibration sources for UV and X-ray spectrometers. This is because
their spectra are simple and dominated by hydrogen features. In some cases there
are small traces of He visible in the spectrum. The simplicity of their
constitution and in particular for the hotter stars the straightforward physics
of the atmosphere allows ab initio calculations of the emerging spectra. The
models depend only upon a few free parameters, predominantly the effective
temperature $T_{\mathrm{eff}}$, the surface gravity $g$ and the relative helium
to hydrogen number density. The effective temperature and surface gravity can be
deduced from optical or UV spectra (Balmer and Lyman series). Despite the
relatively low surface temperatures, that are generally below $10^5$~K, DA white
dwarfs emit soft X-rays because of the low opacity of the atmosphere for the
high energy photons that are produced in the deep and hot inner layers of the
star \citep{shipman1976}. Therefore, if an independent estimate of the radius of
the star as well as its distance are known, the ab initio calculations yield in
principle a model spectrum for the full spectral range including the X-ray band,
with known shape and normalisation (excluding ISM opacity). By comparing such a
model spectrum to the X-ray spectrum as observed with a given instrument, the
effective area of that instrument can be calibrated.

Even when the distance or radius of the star are not accurately known,
measurements of the absolute optical flux of the star can be used to scale the
X-ray flux of the star. In the optical band, absolute fluxes can be determined
with an accuracy of the order of a percent \citep{holberg2006}.

The Chandra low energy transmission grating spec\-tro\-meter (LETGS) is
currently the most sensitive high-resolution X-ray spectrometer operating in
the softest X-ray band. The effective area calibration as produced by the
Chandra X-ray Center was described by \citet{pease2000}; essentially, a pure
hydrogen, non-LTE model for Sirius~B with $T_{\mathrm{eff}}=25\,000$~K and
$g=10^7$~m\,s$^{-2}$ was used \citep{holberg1998}, with an estimated flux
uncertainty of less than 10~\%. A correction factor to the effective area was
determined by comparing the observed spectrum of Sirius~B to this model. Using
the resulting corrected effective area, the observed LETGS spectrum of HZ~43A
agreed within 10--15~\% with a model spectrum for that source. Later
refinements of the effective area \citep{pease2003} mainly concerned the
mid-energy range, based on spectra of the blazars \object{PKS~2155-304},
\object{3C~273} and the isolated neutron star \object{RXJ~1856.5$-$3754}.

Alternatively, at SRON, J. Kaastra and J. Heise derived the effective area of
the LETGS following a different approach described in an internal report
\citep{kaastra2000}. Basically, a grid of models was calculated using a version
of Tlusty \citep{hubeny1988,hubeny1995} available at that time, with the
effective temperature and gravity as free parameter, for both Sirius~B and
HZ~43A. A 6-dimensional grid search was done (varying for each source the
effective temperature, gravity and interstellar absorption column), in order to
find the best matching spectrum using as constraints {\sl (i)} that the ratio of
both model spectra should match as closely as possible the measured count ratio
with the LETGS; {\sl (ii)} that the 70--170~\AA\ spectrum of Sirius~B must be
within the error bars of the model as derived by \citet{holberg1998}; and {\sl
(iii)} that the model parameters should not be too far off from the parameters
for both stars as listed in the literature. 

The effective area of the LETGS as derived by \citet{kaastra2000} and 
\citet{pease2000} differ typically by 10~\%, but the differences are not
constant as a function of wavelength.

More recently, \citet{beuermann2006} have tackled the problem of
cross-calibration again using LETGS spectra of Sirius~B, HZ~43A as well as
RXJ~1856.5$-$3754. Assuming a double blackbody model for the latter source, and
running grids of white dwarf models for the first two sources, they derived the
effective area of the LETGS by fitting simultaneously the parameters of these
stars and the effective area correction factor. Based on their solution, they
concluded that the Extreme UltraViolet Explorer (EUVE) short wavelength (SW)
detector effective area is off by $\sim 15$~\% and that the ROSAT PSPC detector
agrees within a few percent with the LETGS.

However, the assumption of a double blackbody model for the neutron star is not
unquestionable. For instance, the source could have a more complicated
atmosphere, or there may be multiple hot spots or a single spot with
multi-temperature structure. Also, comparing our model calculations to the
models derived by \citet{beuermann2006} shows some significant differences, even
if the same stellar parameters are used (see Sect.~\ref{sect:comparebeuermann}).
This gives us sufficient motivation to re-examine carefully the calibration of
the LETGS and the model spectra employed for both white dwarfs. 

This paper is the second of a series of three intended to derive the absolute
effective area of the reflection grating spectrometer (RGS) of XMM-Newton and
through this of other instruments.  In our first paper \citep{paper1} we have
studied the RGS spectrum of \object{Crab} and derived accurately the
interstellar absorption towards that source. However, the main uncertainty of 10
percent on the absolute flux of Crab could not be resolved. In the present paper
we show how we can accurately calibrate the low-energy effective area of the
Chandra LETGS using white dwarfs. In our third paper \citep{paper3} we combine
these results on white dwarfs with the Crab results using blazar spectra taken
simultaneously with the RGS and LETGS. This leads to a reduction of the
uncertainty in the Crab flux and absolute effective area of the RGS to about 3
percent.

\section{Data analysis and spectral modelling\label{sect:datamodel}}

We follow the same procedures as \citet{kaastra2000} but use the currently best
available spectral models and boundary conditions. In Sect.~\ref{sect:letgs} we
derive the intensity ratio of the spectra of Sirius~B and HZ~43A as measured by
the Chandra LETGS. We will adjust the parameters of both stars using a number of
boundary conditions until the best agreement with this observed intensity ratio
is obtained. Sect.~\ref{sect:wdmodel} describes our model for the white dwarf
atmospheres. These model spectra essentially give the surface flux of the
atmosphere, so we describe the scaling procedure to obtain the flux received at
Earth in Sect.~\ref{sect:scaling}. In Sect.~\ref{sect:constraints} we list all
constraints that we apply to our models, and Sect.~\ref{sect:fitting} describes
our spectral fitting procedure.

\subsection{Data analysis\label{sect:letgs}}

\begin{table*}[!ht]
\caption{Summary of LETGS observations used in this paper.}
\smallskip
\label{tab:obslog}
\centerline{ 
\begin{tabular}{lcccc}
\hline\hline
Source                & Sirius~B         & Sirius~B         & Sirius~B  & HZ~43A \\
ObsID                 & 1421             & 1452             & 1459      & 59 \\
\hline
start date            & 28 Oct 1999      & 26 Oct 1999      & 27 Oct 1999      & 12 Nov 1999 \\
Duration (ks)         & 25.3             & 28.0             & 12.0             & 40.2 \\
Net exposure (ks)$^a$ & 13.8             & 19.0             & 11.8             & 38.9 \\
0th order counts/s    & $1.282\pm 0.010$ & $1.249\pm 0.008$ & $1.310\pm 0.011$ & $8.826\pm 0.015$ \\
PHA difference$^b$    & $-5.6\pm 2.0$    & $-16.2\pm 1.9$   & $-6.5\pm 2.0$    & $-7.0\pm 1.4$ \\
\hline
\end{tabular}
}
\smallskip
\begin{list}{}{}
\item[${\mathrm{a}}$] Dead-time corrected exposure time after filtering out periods of high
background.
\item[${\mathrm{b}}$] The PHA (pulseheight) difference is the average PHA of the spectrum
relative to a standard spectrum.
\end{list}
\end{table*}

The data used in this paper are summarised in Table~\ref{tab:obslog}. All data
were obtained with the high resolution camera (HRC-S) in combination with the
low energy transmission grating (LETG). We will call this combination here
LETGS, as in our paper we do not use the LETG combined with the ACIS-S detector.

For Sirius~B we used all three LETGS spectra taken October 26--29, 1999. The net
total exposure time is 44.6~ks.  Recently (18 January 2008) the source was
observed again for calibration purposes but as the source was close to the
detector edge we do not use these data. For HZ~43A we used observation ID 59,
observed November 12, 1999, with a net exposure time of 38.9~ks. HZ~43A is
monitored regularly since then, but as we are only interested in the ratio of
the spectra of Sirius~B to HZ~43A, and HZ~43~A is the strongest source, the
statistical uncertainty on the ratio is dominated by the statistical uncertainty
of the Sirius~B spectrum. Moreover, the sensitivity of the LETGS slowly
decreases over time \citep{beuermann2006}, and therefore it is important to
compare observations not too far apart in time.

The data processing is similar to \citet{kaastra2002}. Periods with high
background are filtered out in two steps. In the first step, the observation is
split into 1~s intervals; all intervals where the total detector count rate
saturates (more than 180 counts\,s$^{-1}$) are discarded. After this a light
curve for the zeroth order spectrum of the source is created, binned in 100~s
intervals. This serves to check the constancy of the source. All intervals with
less than 50~s exposure are discarded. The remaining exposure time is then
corrected for dead time using the standard dead time correction factors taken
every 2.05~s, provided with each dataset. 

A few other health checks are made; in one of these, we compare the average
pulseheight (PHA) of the dispersed spectrum with a reference spectrum. This
comparison is done in $2-20$~\AA\ wide bins, and the average PHA difference of
the spectrum compared to the standard is calculated (see
Table~\ref{tab:obslog}). Clearly, the average PHA in Sirius~B observation 1452
is significantly lower than for all other observations. Also, the average zeroth
order count rate is lower: while the count rates for observation 1421 and 1459
are consistent with an average value of $1.295\pm 0.007$~counts\,s$^{-1}$, for
observation 1452 we have $1.249\pm 0.008$~counts\,s$^{-1}$, or 3.6~\% less
counts. A smaller effect is present in the dispersed spectra of Sirius~B. We
have fitted all three spectra individually and found that their shapes are --
within the error bars -- consistent with the same model, but the average flux
for observation 1452 is $0.8\pm 0.4$~\% smaller than for the other observations.
As observation 1452 contributes 43~\% to the total exposure time for Sirius~B,
we kept the data but corrected the fluxes of the total spectrum by $+0.34$~\% to
account for the lower sensitivity during observation 1452. In this way, the data
for both stars are all corrected to a similar PHA-level. 

The spectra were extracted from a region with half-width $h$ in the cross-dispersion
given by $h=\max(1.02,0.027\vert \lambda\vert)$ where $h$ is in arcsec and
$\lambda$ is the wavelength in \AA. The background was extracted from the two
rectangular regions between 10\arcsec $-$ 40\arcsec\ above and below the source
spectrum, in the cross-dispersion direction. 

\begin{figure}
\resizebox{\hsize}{!}{\includegraphics[angle=-90]{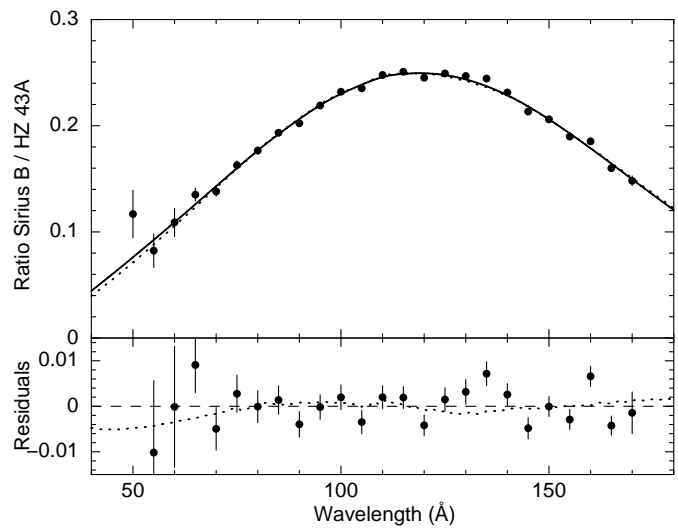}}
\caption{Ratio of the spectrum of Sirius~B to the spectrum of
HZ~43A. data points with error bars: ratio as observed by the LETGS. 
The solid line shows our best-fit model 2 (with a long wavelength
cut-off of the Lyman pseudo-continuum), and the dotted line our best-fit
model 1 (short cut-off), as discussed in Sect.~\ref{sect:results}.
The lower panel shows the fit residuals of the observed ratio with
respect to model 2; the dotted line in that panel shows on the same scale
the differences between model 2 and model 1. Note that the first data point
at 50~\AA\ is off-scale in this lower panel.}
\label{fig:ratcomp}
\end{figure}

The spectra of both stars were fitted with a spline continuum with knots
with a spacing of 5~\AA; the values of the spline at each knot and the
associated uncertainties were determined. This model was then folded through the
response matrix, and higher spectral order contamination, although very small,
was taken into account (see Sect.~\ref{sect:extraction} for an estimate of the
associated uncertainty). From these fits to the count spectra, the ratio between
both photon spectra was calculated. Thanks to the high spectral resolution, this
ratio is independent on any errors in the effective area. The ratio is shown in
Fig.~\ref{fig:ratcomp}.

\subsection{White dwarf model spectra\label{sect:wdmodel}}

The white dwarf models that have been successfully used in modelling the 
spectra of Sirius~B and HZ~43A can be classified as
follows:
\begin{enumerate}
\item Pure hydrogen models
\item Homogeneous H-He models
\item Stratified models with a hydrogen layer atop of a helium atmosphere
\end{enumerate}

We have calculated a set of spectra for these models with a range of parameters
that are appropriate for Sirius~B and HZ~43A, respectively. The NLTE model
atmosphere calculations were done with the latest version (v.~202) of Tlusty
\citep{hubeny1995}. The plane-parallel model atmospheres were discretised with
100 depth points which span a wide range in Rosseland optical depth from
10$^{-5}$ to 10$^4$. The whole spectrum is sampled with about 7\,000 frequencies
between $4\times 10^{17}$ and $10^{12}$~Hz. Hydrogen was represented with a
16-level model atom supplemented with a higher superlevel combining all higher
levels up to $n = 80$. Treatment of level dissolution, pseudo-continuum,
superlevel, and line opacity follows \citet{hubeny1994}. Model atoms of
\ion{He}{i} and \ion{He}{ii} incorporate 24 and 20 levels, up to $n$ = 8 and 20,
respectively; a description of the helium atomic data can be found in
\citet{lanz2003}.

In most cases the differences between the models in the UV and optical part of
the spectrum are small, which is no surprise as the range of parameters has
been adopted from previous studies that estimated these parameters by fitting
the optical and UV spectra. However, as the X-ray spectrum can be considered
more or less as an exponential tail to the full spectrum, a tiny shift in
effective temperature or surface gravity can have a large effect to the X-ray
spectrum. For this reason the X-ray spectra need to be calculated with care.

\subsubsection{Compton scattering}

A potential important effect for the radiative transfer of X-ray photons in
the white dwarf atmosphere is Compton scattering. This was first indicated by
\citet{madej1998} for the case of HZ~43A. Hard X-ray photons scatter many
times before they reach the surface of the star, and the combined effect of
all the collisions with relatively cool electrons can cause a significant
energy loss for these photons, and therefore a softening of the high-energy
tail of the spectrum. This effect is most noticeable at wavelengths smaller
than 100~\AA. However, recently \citet{suleimanov2006} showed using a more
sophisticated calculation that for both HZ~43A and Sirius~B Compton scattering
can be completely ignored. Therefore, in the present calculations we will also
ignore Compton scattering.

\subsubsection{Hydrogen Lyman pseudo-continuum opacity\label{sect:pseudo}}

When we compared our calculations in detail with the calculations of
\citet{beuermann2006} (see Sect.~\ref{sect:comparebeuermann} for more details),
we got a good agreement for HZ~43A but large differences for Sirius~B in the
EUV/X-ray band. A detailed investigation of the problem showed that the
differences are due to the treatment of the Lyman pseudo-continuum.

What basically happens is the following. In the high density atmosphere of
Sirius~B, the density is high enough that a fraction of the atoms have
neighbours so nearby that they are significantly perturbed.  The highest energy
levels for these atoms become partly dissolved, that is an atom ending in them
can be viewed as ionised, and these levels are viewed as partly dissolved, and
partly truly occupied. \citet{hummer1988} have considered this phenomenon in
detail,  and calculated well-defined occupation probabilities for hydrogenic
levels.

Consequently, in addition to a traditional continuum that extends from the
photoionisation threshold shortward and that corresponds to a true
photoionisation, there is also a "pseudo-continuum" that extends longward of the
threshold, and that corresponds to transitions from a bound lower level to the
dissolved parts of  higher levels. The basic physical process is well-defined,
because a transition to the dissolved part of the level leaves an atom indeed in
an unbound state and thus the process is a sort of photoionisation, but the
essential problem is how to formulate the appropriate cross-section for the
pseudo-continuum. According to the standard approach
\citep{dappen1987,hubeny1994}, the cross-section is formulated through an 
extrapolation of the traditional photoionisation cross-section shortward of the
edge, and the so-called "dissolved fraction" that is given through a dissolution
probability of a fictitious level that would correspond to the current frequency
$\nu$ (for details, refer to the above cited papers). \citet{hubeny1994}
outlined a proof of this assertion, but stressed that the proof only applies
"close to the photoionisation limit". It is hard to specify exactly how far from
the edge is the formalism valid, but it was certainly never meant to extend
hundreds or even thousands of {\AA}ngstroms away from the edge. Therefore, most
researchers begun to use ad hoc cut-off frequencies for the pseudo-continuum
cross-sections, with the belief that their actual values do not significantly
influence  modelling results. Some researchers have even introduced smoothed
cut-off regions (Bergeron, private communication), but in any case it should be
clearly understood that any cut-off is a completely ad hoc concept.

In fact, a more rigorous treatment of this problem should be developed, perhaps
analogously to a recent treatment of the pseudo-continuum cross-sections for
neutral perturbers by \citet{kowalski-saumon2006} and \citet{kowalski2006}, but
this was not yet done. For the moment, we are left with a necessity of using
artificial cut-offs. The TMAP code \citep{werner2003} employed by
\citet{beuermann2006} uses a long cut-off at 2431~\AA, while the default for
Tlusty, employed in the present work, is a short cut-off at 925~\AA, close to
the Lyman edge. Lacking a better physical model for the pseudo-continuum, it is
hard to decide which approach is better. Therefore we use in this paper two
different sets of calculations: one with a short cut-off (referred to here as
model 1), the other with a long cut-off (model 2).

\subsection{Scaling of the white dwarf spectra\label{sect:scaling}}

The model spectra calculated with our code gives the Eddington flux $H_\nu$
(usually expressed in units equivalent to W\,m$^{-2}$\,Hz$^{-1}$). The photon
spectrum $N_\lambda$ (in photons\,m$^{-2}$\,s$^{-1}$\,\AA$^{-1}$) seen at Earth
is then given by

\begin{equation}
N_{\lambda} = {f_d} {4\pi  H_\nu \over  h\lambda} T(\lambda),
\label{eqn:spectrum}
\end{equation}

where $T(\lambda)$ is the transmission of the ISM and $f_d \equiv R^2/d^2$ with
$R$ the radius of the star and $d$ the distance to the star.

For Sirius~B, the distance is known accurately with a precision of 0.4~\% (the
HIPPARCOS-based parallax is $0\arcsec .3792\pm 0\arcsec .0016$,
\citealp{holberg1998}). Its radius $R$ can be determined in principle from the
surface gravity $g=GM_{\mathrm B}/R^2$, since the white dwarf mass $M_{\mathrm
B}$ is known with an accuracy of 1.5~\%. However, the typical uncertainty in
$g$ is 15~\% based on fits to the optical and (extreme) ultraviolet spectra
\cite[for example][]{holberg1998}. Using the currently best available value
for the gravitational redshift (see later), the uncertainty in $g$ is still
12~\%. Therefore, using this scaling in the form $f_d = GM_{\mathrm B}/ gd^2$
gives a flux uncertainty of at least 12~\%.

There is a simple way out, however. We define $f^{\prime} \equiv f_d g$ and
substitute $f_d = f^{\prime} / g$ in (\ref{eqn:spectrum}). For a given
spectral model, $g$ is prescribed as one of the input parameters of the model
and therefore exactly known. It is easy to show that
\begin{equation}
f^{\prime} = \frac{4\pi^2 a_p^3 d}{P^2\mu_{\mathrm B}^2},
\label{eqn:scaling}
\end{equation}
with $a_p = 2\arcsec .490 \pm 0\arcsec .004$ the photocentric semi-major axis
\citep{gatewood1978}, $P = 50.09 \pm 0.06$~yr the orbital period
\citep{vandenbos1960}, and $\mu_{\mathrm B} = M_{\mathrm B} / (M_{\mathrm A} +
M_{\mathrm B}) = 0.3295\pm 0.0010$ the relative mass of Sirius~B
\citep{gatewood1978}. Substituting these numbers, we obtain a value of 
$f^{\prime} = (2.084\pm 0.019)\times 10^{-14}$~m\,s$^{-2}$, implying that we
can determine -- given the surface gravity corresponding to the model being
considered -- the flux with an accuracy of 0.9~\%.

For HZ~43B, the distance is less accurately known, and we use here the observed
flux in the optical band to scale the spectrum. The most accurate and best
calibrated flux is obtained in the V band: the V magnitude of HZ~43 is
$12.909\pm 0.0017$ \citep{bohlin2000}. \citet{holberg2006} give the relation
between the monochromatic flux $F_\lambda$ at 5423~\AA\ and the V magnitude as
$F_\lambda = 3.804\times 10^{-9} \times 10^{-0.4V}$, within about 0.5~\%. This
corresponds to a photon flux of 71.25 phot\,m$^{-2}$\,s$^{-1}$\,\AA$^{-1}$ at
5423~\AA. We use this flux to scale all our model calculations, and assign a
nominal uncertainty of 1~\% to it. Note that a similar procedure is harder to
apply to Sirius~B, as contamination by Sirius~A enhances photometric
uncertainties (see Sect.~\ref{sect:constraints}).

\subsection{Constraints on the spectral models\label{sect:constraints}}

\subsubsection{Gravitational redshift\label{gravred}}

The gravitational redshift of Sirius~B is now known much more accurately using
STIS data \citep[$\varv_g = 80 \pm 5$~km\,s$^{-1}$,][]{barstow2005} then
previously (for example $89 \pm 16$~km\,s$^{-1}$, \citealt{greenstein1971};
$85 \pm 15$~km\,s$^{-1}$, \citealt{hebrard1999}). This more accurate
gravitational redshift, combined with the mass derived from astrometric
methods allows an independent estimate of the surface gravity
$g=\varv_g^2c^2/GM_B$ or
\begin{equation} 
g = \frac{\varv_g^2 c^2 P^2 \mu_{\mathrm B}^2}{4\pi^2a_p^3 d^3},
\end{equation}
using the same fundamental parameters as employed for (\ref{eqn:scaling}). This
leads to $\log g$~(m\,s$^{-2}$)~$= 6.62\pm 0.05$. In our spectral modelling, we
have discarded any solution where $\log g$ deviates more than $2\sigma$ from the
above value.

For HZ~43A, \citet{reid1996} gives a value of $\varv_g=30.1$~km\,s$^{-1}$;
\citet{kruk2002} estimate the accuracy of this value to be 10--15~km\,s$^{-1}$.
Therefore we limit our models to those cases where the gravitational redshift of
HZ~43A is within the 10--50~km\,s$^{-1}$ range. We do this as follows. The
parallax of HZ~43~A is known to be $15.3\pm 2.9$~mas \citep{vanaltena1995}. We
use this to derive the minimum and maximum allowed distance $d$. For a given
surface gravity, this distance range corresponds to an allowed range for
$\varv_g=g d \sqrt{f_d} /c$ when we use the accurately known value of
$f_d=R^2/d^2$ derived from the optical flux (Sect.~\ref{sect:scaling}). We then
add this uncertainty in quadrature to a nominal uncertainty of 15~km\,s$^{-1}$
in the observed redshift. Solutions off by more than $2\sigma$ are discarded.
Also whenever the predicted range for $\varv_g$ does not overlap with the
10--50~km\,s$^{-1}$ range, we discard the solution.

\subsubsection{Optical and UV flux of Sirius~B\label{sect:sirfluxes}}

For HZ~43A, we already scale our spectra to agree with the optical flux of this
star. Our model spectra for Sirius~B must also be in agreement with optical and
UV flux measurements (Table~\ref{tab:siriusflux}). We only consider here
monochromatic fluxes, and avoid the use of magnitudes as this involves an extra
complication, namely the convolution with filter transmissions. Care should be
taken in assessing the uncertainties in those flux measurements.

\citet{hebrard1999} analysed the HST-GHRS Echelle-A spectrum of Sirius~B. From
their Fig.~2 we measure the continuum flux at 1302~\AA\ as $1.270 \times
10^{-13}$ W\,m$^{-2}$\,s$^{-1}$\,\AA$^{-1}$ or $(8.32\pm 0.07) \times 10^4$
photons\,m$^{-2}$\,s$^{-1}$\,\AA$^{-1}$. Although the statistical errors are
small, systematic effects are larger as shown below. 

\citet{mack1997} have calibrated the GHRS Echelle-A spectrometer by comparing
spectra with IUE spectra of the standard star $\mu$~Col. From their Fig.~4c  we
find that the relative calibration accuracy of this instrument between
1250--1350~\AA\ is 0.84$\pm$0.36 percent (the residual r.m.s. scatter between
the GHRS and IUE spectra). Thus, the flux depends critically upon the
calibration of IUE.

\citet{bohlin1990} used more than 2700 individual IUE SWP and LWR spectra to
define the absolute flux distributions of the 37 HST standard stars in the
wavelength range 1150--3300~\AA. They conclude that the systematic external
errors in the fluxes are less than 15 percent, while comparison with ANS flux
measurements demonstrates an internal consistency of the IUE spectrophotometry
of 2 percent. The basis for the absolute flux scale in the UV is given by the
spectrum of $\eta$~UMa, which has an uncertainty of 10~\% in its absolute flux
scale according to \citet{bohlin1990}. IUE fluxes may be too low by
$\sim$~10~\%.

In summary, we combine the following errors: systematic uncertainty IUE scale:
10~\%; internal uncertainty IUE specrum $\mu$~Col: 2~\%; relative error
GHRS/IUE: 0.84~\%; statistical error GHRS spectrum Sirius~B: 0.8~\%. The
resulting total uncertainty is 10.3~\%, obviously dominated by the uncertainty
in the absolute UV flux of $\eta$~UMa. We therefore adopt a 1302~\AA\ flux of
$8.32\pm 0.86 \times 10^4$ photons\,m$^{-2}$\,s$^{-1}$\,\AA$^{-1}$. For this
wavelength, we can neglect any ISM continuum extinction.

\begin{table}[!h]
\caption{Measured monochromatic fluxes of Sirius~B. 
}
\label{tab:siriusflux}
\centerline{ 
\begin{tabular}{rcccr}
\hline\hline
Wavelength &  flux  & stat. error & syst. error & ref.\\
(\AA)      & (ph\,m$^{-2}$\,s$^{-1}$\,\AA$^{-1}$) & (\%) & (\%) & \\
\hline
 250       & $19.5$            & 10  & $\pm$10     & a \\
 700       & $20$              & -   & $\pm$100    & b \\
1302       & $8.32\times 10^4$ & 0.8 & $\pm$10.2   & c \\
3500       & $9.41\times 10^4$ & 1.0 & $+$4.5      & d \\
4600       & $6.62\times 10^3$ & 1.0 & $+$4.5      & d \\
6400       & $2.64\times 10^3$ & 1.0 & $+$4.5      & e \\
6700       & $2.25\times 10^3$ & 1.0 & $+$4.5      & e \\
\hline
\end{tabular}
}
\smallskip
\begin{list}{}{}
\item[${\mathrm{a}}$] EUVE MW, \citet{holberg1998}
\item[${\mathrm{b}}$] EUVE LW, \citet{craig1997}
\item[${\mathrm{c}}$] HST-GHRS, \citet{hebrard1999}
\item[${\mathrm{d}}$] HST-STIS G430L, \citet{barstow2005}
\item[${\mathrm{e}}$] HST-STIS G750M, \citet{barstow2005}
\end{list}
\end{table}

\citet{barstow2005} have analysed STIS spectra of Sirius~B. From their Fig.~5,
we estimate the flux at 3500 and 4600~\AA\ (using the G430L grating), and from
their Fig.~4, we estimate the flux at 6400 and 6700~\AA\ (using the G750M
grating).  The typical statistical uncertainty combined with the uncertainty on
the STIS flux scale is about 1~\%, and we adopt that as the nominal statistical
uncertainty. \citet{barstow2005} argue that although STIS has been calibrated to
a nominal precision of $\sim 1$~\%, the necessary usage of the narrow
52\arcsec$\times$0.2\arcsec\ slit for Sirius~B gives an additional 4.5~\% error,
as estimated first by \citet{bohlin1998}. This error, however, always leads to
an underestimate of the flux. Therefore, we add a one-sided, positive systematic
uncertainty of 4.5~\% to the fluxes.

In practice, for the five UV and optical flux values we add the statistical and
systematic uncertainties in quadrature, and calculate for each point the
contribution to a formal $\chi^2$ when comparing to a set of model fluxes. In
this, the one-sidedness of the systematic uncertainties in STIS fluxes is taken
into account. Whenever this $\chi^2$ (with 5 degrees of freedom) exceeds the
$2\sigma$ upper limit (at $\chi^2=5+2\sqrt{10}$) we discard the solution.

\subsubsection{EUVE flux of Sirius~B\label{sect:euve_sir}}

The EUV flux of Sirius~B is a very sensitive indicator for the spectral
parameters of this source \citep{holberg1998}. Unfortunately, there is little
information publicly available about the effective area calibration of EUVE. 

The following information is given in the EUVE guest observer
handbook\footnote{available at
http://archive.stsci.edu/euve/handbook/handbook.html}. It explains at page 2--15
the calibration procedure, paraphrased below. "Because of the lack of standard
stars in the EUV, the effective area of the spectrometers was determined from
model spectra for continuum sources. White dwarf stars provided the baseline
measurements. After spectra were extracted, white dwarf model spectra for the
object were input to a spectrometer simulation program, and the model parameters
adjusted to produce simulated data that match the observed spectra as nearly as
possible. In most cases, $T_{\mathrm{eff}}$ and $g$ were well constrained by
other optical and UV spectra. The resulting models were then compared whenever
possible to photometric and spectroscopic EUV/soft X-ray observations of the
same target made with the ROSAT WFC, EXOSAT, HUT, IUE, and various rocket
experiments. The effective area function was then readjusted to bring the
measurements into better agreement."

However, given the uncertainty of white dwarf fluxes measured with previous
instruments in both the UV and soft X-ray band, as well as the strong dependence
of the soft X-ray flux on slight changes in the spectral parameters, we believe
there is significant systematic uncertainty involved; moreover, the ground
calibration was accurate to about 25~\%. A fit to the EUVE spectrum of HZ~43 by
\citet{barstow1995} shows remaining residuals of the order of 5~\%. Also the
analysis of \citet{sing2002} of a sample of 7 DA white dwarfs shows systematic
residuals up to 10~\%, with a typical scatter of 5~\%. We therefore
assume that the absolute fluxes measured with EUVE have a systematic uncertainty
of 10~\%.

We constrain our models using the EUVE flux measurement at 250~\AA\ 
($1.55\times 10^{-16}$~W\,m$^{-2}$\,\AA$^{-1}$), 
taken from Fig.~1 of \citet{holberg1998},
with a 10~\% statistical error margin added in quadrature to the systematic
uncertainty, to account for both the statistical and systematic uncertainties in
the EUVE spectrum. When the EUVE flux point at 250~\AA\ exceeds the 1$\sigma$
bound, we discard the solution.

Finally, we list in Table~\ref{tab:siriusflux} a constraint from the longest
wavelength EUVE LW spectrometer, although the error on this estimate is large.
The main reason for inclusion is that the spectrum at this wavelength is
rather sensitive to the foreground absorption model, and we want our model not
to be in conflict with this EUV constraint. 

\subsubsection{Interstellar absorption\label{sect:ism}}

\begin{table}[!h]
\caption{Adopted absorption parameters for Sirius~B and HZ~43A.
Numbers in brackets are derived from the basic numbers given in the other
entries of this table.} 
\smallskip
\label{tab:abspar}
\centerline{ 
\begin{tabular}{lcc}
\hline\hline
Ion       & Sirius~B & HZ~43A  \\
\hline
\ion{H}{i}   column ($10^{22}$~m$^{-2}$)  & 0.58 $\pm$0.11  & 0.85 $\pm$0.06   \\
\ion{He}{i}  column ($10^{22}$~m$^{-2}$)  &(0.045$\pm$0.010)& 0.057$\pm$0.002  \\
\ion{He}{ii} column ($10^{22}$~m$^{-2}$)  &(0.017$\pm$0.004)& 0.039$\pm$0.012  \\
\ion{H}{i} / \ion{He}{i} &  12.8$\pm$1.4  &(14.9$\pm$1.2)\\
\ion{He}{ii} / (\ion{He}{i}+\ion{He}{ii}) &0.27$\pm$0.04 &(0.41$\pm$0.08) \\
\hline
\end{tabular}
}
\smallskip
\begin{list}{}{}
\item[${\mathrm{a}}$] EUVE MW, \citet{holberg1998}
\item[${\mathrm{b}}$] EUVE LW, \citet{craig1997}
\item[${\mathrm{c}}$] HST-GHRS, \citet{hebrard1999}
\item[${\mathrm{d}}$] HST-STIS G430L, \citet{barstow2005}
\item[${\mathrm{e}}$] HST-STIS G750M, \citet{barstow2005}
\end{list}
\end{table}

The adopted values of the interstellar absorption towards Sirius~B and HZ~43A
are shown in Table~\ref{tab:abspar}. The hydrogen column towards Sirius~B is the
weighted average of $(0.65\pm 0.16)\times 10^{22}$~m$^{-2}$ \citep[][from
Ly$\alpha$ with the HST-GHRS detector]{hebrard1999} and  $(0.52\pm 0.14)\times
10^{22}$~m$^{-2}$ \citep[][from Ly$\alpha$ with IUE detector]{holberg1998}.
Unfortunately Sirius~B has a low EUV flux such that only upper limits to the He
column densities are known: $<0.03\times 10^{22}$~m$^{-2}$ for \ion{He}{i} and
$<0.14\times 10^{22}$~m$^{-2}$ for \ion{He}{ii} \citep{wolff1999}. Therefore, we
adopt the average \ion{H}{i}/\ion{He}{i} ratio of $12.8\pm 1.4$ derived for the
local environment of the Sun \citep{slavin2007}. Similarly, we follow
\citet{holberg1998} who use the average \ion{He}{ii}/He ratio of $0.27\pm0.04$
derived by \citet{barstow1997}. Note that the uncertainty in the hydrogen column
of Sirius~B leads to an uncertainty of 34~\% in the predicted flux at 700~\AA.
At shorther wavelengths the corresponding flux uncertainty is smaller, for
example only 2~\% at the peak of the soft X-ray spectrum (150~\AA).

The neutral hydrogen column towards HZ~43A (Table~\ref{tab:abspar}) has been
discussed in detail by \citet{kruk2002}, and we use their value. For \ion{He}{i}
and \ion{He}{ii} we use the weighted average of 
\citet{dupuis1995,wolff1999,barstow1997}, based on EUVE measurements. 

Due to the low column densities, other ions than those from hydrogen and helium
can be ignored in the Chandra band: we estimate that between 1--180~\AA\ metals
contribute less than 0.3~\% to the continuum opacity. For our opacity
calculations, we use the model by \citet{rumph1994}. The uncertainty in the
measured column densities hardly affects our estimated flux of HZ~43A in the
Chandra band: at 180~\AA\, where the ISM opacity is highest, the uncertainty in
the column densities of \ion{H}{i}, \ion{He}{i} and \ion{He}{ii} corresponds to
a transmission uncertainty of 0.3, 0.2 and 1.0~\%, respectively. The combined
r.m.s. error is 1.1~\%. Hence, the uncertainty in our model is mainly determined
by the intrinsic parameters of HZ~43A itself.

In our spectral modelling, for both stars we allowed the column densities for
HZ~43A and Sirius~B (only \ion{H}{i}) or the ratios \ion{H}{i}/\ion{He}{i} and
\ion{He}{ii}/He (for Sirius~B) to vary between the $\pm 2\sigma$ limits as given
in Table~\ref{tab:abspar}.

\subsection{Parameter estimation\label{sect:fitting}}

Our model depends on twelve parameters, six for each star, namely
the effective temperature, surface gravity, helium abundance or hydrogen layer
thickness, and interstellar column densities of \ion{H}{i}, \ion{He}{i} and
\ion{He}{ii}. Often there are strong correlations
between these parameters, and we have a number of constraints to be obeyed
(Sect.~\ref{sect:constraints}). In summary, these constraints are:
\begin{enumerate}
\item 250~\AA\ EUVE flux of Sirius~B 
\item Optical and UV fluxes of Sirius~B at 1302, 4600, 6400 and 6700~\AA\
\item $\log g$ for Sirius~B within the 2$\sigma$ range of the value
derived from the gravitational redshift
\item Gravitational redshift of HZ~43A
\item derived column densities (or ratios) towards both stars within
their 2$\sigma$ error limits.
\end{enumerate}

We define as usual a quantity $\chi^2$ given by
\begin{equation}
\chi^2 = \sum_i (O_i - M_i)^2 / \Delta O_i^2
\label{eqn:chisq}
\end{equation}
where $O_i$ and $\Delta O_i$ is the observed LETGS ratio and the associated
uncertainty of the Sirius~B to the HZ~43A spectrum, and $M_i$ is the predicted
ratio based on our model. We use the data points between 50--175~\AA\ with a
spacing of 5~\AA\ as derived in Sect.~\ref{sect:letgs}. Whenever any of the
constraints of Sect.~\ref{sect:constraints} is violated, we formally add to
$\chi^2$ a large number (1000) in order to discard that solution. However,
as it is more likely that our constraints are near the expected value
than at their extremes, we add for each of the above five constraints
a nominal $\Delta\chi^2$ to (\ref{eqn:chisq}) corresponding to the number
of standard deviations for that constraint.

We find the best solution using a Monte Carlo method. Starting with a broad
range of allowed parameters, we draw random sets of parameters within that
allowed range, and evaluate $\chi^2$ for each set. Solutions with $\chi^2$
larger than a threshold are discarded. After having obtained a sufficient number
of solutions, we slowly decrease the $\chi^2$ threshold and simultaneously
shrink the allowed parameter space, encompassing with some margin all solutions
that up to then have been acceptable. All acceptable solutions are stored, and
after having reached the best solution with $\chi^2=\chi^2_{\min}$ we find the
errors on the parameters by finding for each parameter the minimum and maximum
value for which $\chi^2<\chi^2_{\min}+1$. We also store each acceptable
spectrum, so we can also determine for each wavelength the range of allowed flux
values.

\section{Models to be used for Sirius~B and HZ~43A}

\subsection{Sirius~B}

For Sirius~B we have used homogeneous models, which include a pure hydrogen
atmosphere as limiting case.

We have also calculated a grid of stratified models for Sirius~B, but we were
not able to obtain successful fits. Basically, we constrained the photometric
hydrogen column to the range of $(1.00-1.25)\times 10^{-13}$~\Msun, around the
value of $1.13\times 10^{-13}$~\Msun\ found by \citet{holberg1998} for this
class of models. The main reason for the failure is that the stratified models
show a flux deficit of up to a factor of 2--3 around 50~\AA\ as compared to
homogeneous models (see also Fig.~\ref{fig:hlaag}); the deficit sets on below
80~\AA. As this range was at the short wavelength end of the EUVE spectrometer,
\citet{holberg1998} were not able to exclude this class of models completely.
Thanks to the sensitivity of Chandra it is now possible to rule out this class
of models.

\subsection{HZ~43A}

For HZ~43A we first consider the homogeneous models. \citet{barstow1995} have
put strict upper limits to the amount of He in HZ~43A, based on the limits to
the 304~\AA\ line of \ion{He}{ii} in the EUVE spectrum. The nominal equivalent
width of this line derived by Barstow et al. is 0.2$\pm$0.1~\AA, but due to
possible systematic effects in the EUVE spectrum this cannot be regarded as a
detection. For their mixed He/H models, they obtain an upper limit of $3\times
10^{-7}$ for the He/H ratio. We have calculated a grid of homogeneous models
with He/H ratio's between 0 and $10^{-5}$. Our models with a small ratio such
as found by \citet{barstow1995} yield fluxes in the Chandra band (10--180~\AA)
that are 1.4--2.7~\% smaller than the fluxes for a pure H model, for the same
values of $T_{\mathrm{eff}}$ and $g$. It is clear that such small differences
can be easily accommodated for in a pure H model using slightly different
values for $T_{\mathrm{eff}}$ and $g$, which are still consistent with the
limits from other parts of the spectrum to these numbers. We conclude that --
at least for our calibration purposes -- we can safely adopt a pure hydrogen
model as far as the class of homogeneous models is concerned.

\begin{figure}
\resizebox{\hsize}{!}{\includegraphics[angle=-90]{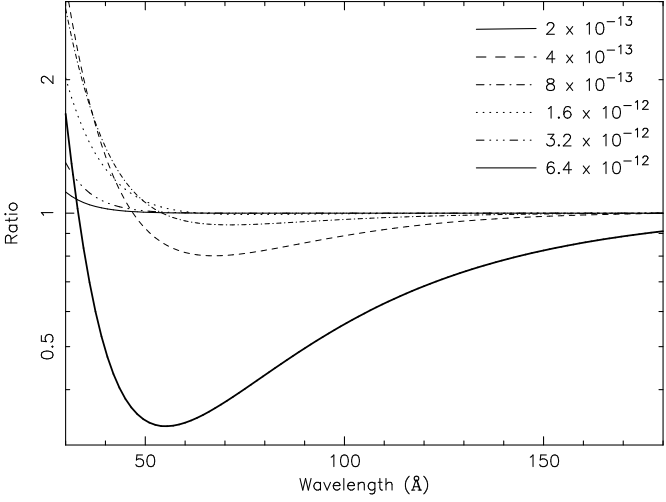}}
\caption{Spectrum of HZ~43 for stratified models with hydrogen layers
(in \Msun) as indicated on the plot, divided by the spectrum of a pure
hydrogen model. The calculation is done for $\log g$ (m\,s$^{-2})= 5.946$ and
$T_{\mathrm{eff}}=50551$~K (the best-fit parameters for model 2, 
see Table~\ref{tab:bestpar}).}
\label{fig:hlaag}
\end{figure}

The other important class of models that include He are the stratified models.
We have made a grid of models with a hydrogen layer mass between $10^{-14}$ and
$10^{-10}$~\Msun. All models with a hydrogen layer less than $10^{-13}$~\Msun\
produce too deep He features in the spectrum, consistent with the findings of
\citet{barstow1995}. On the other hand, if the hydrogen layer mass is
$>10^{-11}$~\Msun, the spectrum cannot be distinguished from a pure hydrogen
atmosphere in the Chandra band. However in the case of an intermediate
thickness hydrogen layer of $\sim 10^{-12}$~\Msun\ there is a significant
effect in the Chandra band (Fig.~\ref{fig:hlaag}). For the lowest allowed
hydrogen column ($2\times 10^{-13}$~\Msun, see below), there is a $\sim$70~\%
flux reduction around 55~\AA, diminishing rapidly for larger columns, while at
shorter wavelengths the flux is higher, peaking for $4\times 10^{-13}$~\Msun\
and then diminishing rapidly in the pure hydrogen limit.

\begin{table}[!h]
\caption{Lyman series of \ion{He}{ii} in HZ~43. 
}
\smallskip
\label{tab:hzhelines}
\centerline{ 
\begin{tabular}{lrr}
\hline\hline
Line & $\lambda$ (\AA) & EW (\AA)  \\
\hline
Ly$\alpha$  & 304 & 0.14$\pm$0.11  \\
Ly$\beta$   & 256 & 0.05$\pm$0.07  \\
Ly$\gamma$  & 243 & 0.09$\pm$0.06  \\
\hline
\end{tabular}
}
\end{table}

For this reason, we carefully reconsider the constraints to the hydrogen layer.
From the EUVE spectrum published by \citet{barstow1995} (their Fig.~2), we
estimated the equivalent widths of the \ion{He}{ii} Ly$\alpha$, $\beta$ and
$\gamma$ absorption lines (Table~\ref{tab:hzhelines}). We compared these
equivalent widths with our model calculations. None of them can be regarded as a
detection. Combining all three lines, we obtain a best fit hydrogen mass
$M=2-4\times 10^{-13}$~\Msun. This mass predicts equivalent widths for the
\ion{He}{ii} Ly$\alpha$, $\beta$ and $\gamma$ lines of 0.02, 0.04 and 0.02~\AA,
respectively. The 99~\% confidence lower limit is 1.4$\times 10^{-13}$~\Msun.
While the lower limit is rather strict, we cannot fully exclude that the
hydrogen layer is very thick (which would make the model effectively a pure
hydrogen model): the pure hydrogen limit is at the 90~\% confidence upper limit.
Another constraint on the thickness of the hydrogen layer is obtained from the
\ion{He}{ii} Lyman limit edge. Effectively, the edge is slightly shifted due to
blending with the higher Lyman series lines and occurs near 230~\AA. We find
from the EUVE spectrum that the edge is invisible, corresponding to an optical
depth of less than 2~\%. This corresponds to a lower limit to the hydrogen mass
of $2.2\times 10^{-13}$~\Msun. 

We conclude that the thickness of the hydrogen layer in HZ~43 is most likely
between 2--4$\times 10^{-13}$~\Msun, with lower values excluded but with
no solid upper limit to the hydrogen mass.

Given all this, we consider only stratified models for HZ~43A with hydrogen
mass $>2\times 10^{-13}$~\Msun; as argued before, a pure hydrogen atmosphere is
a limiting case of this set of models.

\section{Results\label{sect:results}}

Our best fit model 1 (short cut-off) has a $\chi^2$ of $56.17\,(52.50)$, our
best fit model 2 (long cut-off) has $\chi^2 = 49.41\,(46.33)$. The numbers in
brackets denote the contribution of the Chandra data only. With 25 data
points and 12 adjustable parameters the number of degrees of freedom would be
13, and hence the value of $\chi^2$ is slightly enhanced with respect to purely
statistical noise. However, the actual number of degrees of freedom is higher,
as several parameters are strongly correlated and the best fit is rather
insensitive to others (such as the helium abundance in Sirius~B, and in general
the interstellar absorption column densities). Moreover, our additional
constraints also effectively limit the number of degrees of freedom. Although
hard to estimate exactly, the true number of degrees of freedom may be of the
order of 20.

Fig.~\ref{fig:ratcomp} shows the observed ratio of the LETGS spectra of 
Sirius~B to HZ~43A, together with the best fit models 1 and 2. From this figure
it is clear that there is some additional systematic scatter in the data points
(as the models in the $50-170$~\AA\ wavelength range are, as expected, rather
smooth). For instance, the data point at 160~\AA\ deviates by $+3.0\sigma$ or
$+3.7$~\%. In this case, some of the systematic effect may be due to the fact
that this wavelength is close to the edge of the spectrum in the $-1$ spectral
order (the physical edge of the detector). For other data points, the relative
deviations are smaller or less significant. By adding a systematic uncertainty
of only 1 or 2~\% to our ratios, the $\chi^2$ for the best fit model 2 would
reduce from 49.41 to 29 or 16, respectively, i.e. in the acceptable range given
the $\sim$20 degrees of freedom. This reduction by a factor of $\sim$2 in
$\chi^2$ then suggests that we should use $\Delta\chi^2=2.0$ instead of
$\Delta\chi^2=1.0$ for the original fits without systematic uncertainties, in
order to determine the $1\sigma$ confidence limits on the parameters.

\begin{table*}[!htb]
\caption{Best-fit parameters of Sirius~B and HZ~43A.}
\smallskip
\label{tab:bestpar}
\centerline{ 
\begin{tabular}{lc|ccc|ccc}
\hline\hline
Parameter                            &Search range & Best value$^a$ & Median value$^b$ & Allowed range$^c$ & Best value$^a$ & Median value$^b$ & Allowed range$^c$  \\
&&\multicolumn{3}{c|}{Short cut-off Lyman pseudocontinuum} & \multicolumn{3}{c}{Long cut-off Lyman pseudocontinuum} \\
&&\multicolumn{3}{c|}{(model 1)                          } & \multicolumn{3}{c}{(model 2)                         } \\
\hline
HZ~43A:                              &             &            &              &               &            &              &                \\
$T_{\mathrm{eff}}$ (K)               &47000$-$57000& 51660      & 51660        & 51420$-$51880 & 51530      & 51460        & 51240$-$51690  \\
$\log g$ (m\,s$^{-2}$)               &5.8$-$6.3    & 6.064      & 6.065        & 6.031$-$6.102 & 5.893      & 5.913        & 5.872$-$5.941  \\
H-layer mass ($10^{-11}$~\Msun)      &0.02$-\infty$& 5          & 0.4          & 0.3$-\infty$  & 11         & 10           & 0.25$-\infty$  \\
ISM \ion{H}{i}  ($10^{22}$~m$^{-2}$) & 0.73$-$0.97 & 0.85       & 0.85         & 0.81$-$0.90   & 0.85       & 0.84         & 0.79$-$0.91    \\
ISM \ion{He}{i} ($10^{22}$~m$^{-2}$) &0.053$-$0.061& 0.057      & 0.058        & 0.055$-$0.059 & 0.057      & 0.055        & 0.055$-$0.058  \\
ISM \ion{He}{ii} ($10^{22}$~m$^{-2}$)&0.015$-$0.063& 0.038      & 0.042        & 0.029$-$0.047 & 0.038      & 0.042        & 0.029$-$0.048  \\
$f_d=R^2/d^2$ ($10^{-23}$)           &             & 3.037      & 3.037        & 3.025$-$3.050 & 3.038      & 3.041        & 3.031$-$3.053  \\
Grav. redshift (km\,s$^{-1}$)        & 10$-$50     &   45       &   45         & 41$-$50       & 30.0       & 31.5         & 28.6$-$33.6    \\
\hline
Sirius~B:                            &             &            &              &               &            &              &                \\
$T_{\mathrm{eff}}$ (K)               &24000$-$26000& 25360      & 25360        & 25320$-$25410 & 24970      & 24980        & 24940$-$25010  \\
$\log g$ (log m\,s$^{-2}$)           & 6.52$-$6.72 & 6.628      & 6.626        &  6.624$-$6.631& 6.624      & 6.623        &  6.621$-$6.627 \\
He/H ratio ($\times 10^{-6}$)        & 0$-$6       & 0          & 0.2          & 0$-$0.5       & 0          & 0.6          & 0$-$1.0        \\
ISM \ion{H}{i}  ($10^{22}$~m$^{-2}$) & 0.36$-$0.80 & 0.58       & 0.57         & 0.50$-$0.66   & 0.59       & 0.59         & 0.50$-$0.69    \\
ISM \ion{H}{i}/\ion{He}{i}           & 10.0$-$15.6 & 12.8       & 13.0         & 11.8$-$14.0   & 12.8       & 12.4         & 11.9$-$13.9    \\
ISM \ion{He}{ii}/He                  & 0.19$-$0.35 & 0.27       & 0.27         & 0.24$-$0.30   & 0.27       & 0.25         & 0.24$-$0.30    \\
$f_d=R^2/d^2$ ($10^{-21}$)           &             & 4.91       & 4.93         & 4.87$-$4.95   & 4.95       & 4.97         &  4.92$-$4.99   \\
\hline
\end{tabular}
}
\smallskip
\begin{list}{}{}
\item[${\mathrm{a}}$] Best values (minimum $\chi^2$).
\item[${\mathrm{b}}$] Corresponding to the parameters of the
spectrum closest to the median of all allowed spectra.
\item[${\mathrm{c}}$] Corresponding to $\Delta\chi^2=2$ (see text for discussion).
\end{list}
\end{table*}

\begin{table}[!ht]
\caption{Absorbed fluxes (in photons\,m$^{-2}$\,s$^{-1}$\,\AA$^{-1}$) 
of Sirius~B and HZ~43A at selected wavelengths.}
\smallskip
\label{tab:modelfluxes}
\centerline{ 
\begin{tabular}{ccccc}
\hline\hline
 & \multicolumn{2}{c}{Model 1} & \multicolumn{2}{c}{Model 2} \\
$\lambda$ (\AA)& Sirius~B & HZ~43A & Sirius~B & HZ~43A \\
\hline
48   & 0.257 & 3.98 & 0.213 & 3.05\\
50   & 0.416 & 5.83 & 0.346 & 4.54\\
60   & 2.80  & 26.5 & 2.39  & 21.9\\
70   & 10.7  & 75.5 & 9.34  & 65.3\\
80   & 28.3  & 160  & 25.3  & 143 \\
100  & 99.0  & 429  & 92.1  & 400 \\
120  & 192   & 772  & 184   & 738 \\
140  & 253   & 1110 & 246   & 1080\\
160  & 250   & 1400 & 243   & 1360\\
170  & 227   & 1500 & 220   & 1470\\
250  & 19.6  & 1780 & 18.8  & 1760\\
700  & 15.8  & 123  & 14.4  & 123 \\
1302 & 84000 & 2670 & 81000 & 2670\\
3500 & 9570  & 222  & 9540  & 222 \\
4600 & 6580  & 113  & 6590  & 113 \\
5423 & 4330  & 71.3 & 4340  & 71.3\\
6400 & 2720  & 44.5 & 2730  & 44.5\\
6700 & 2400  & 39.0 & 2400  & 39.1\\
\hline
\end{tabular}
}
\end{table}

We list the best-fit parameters in Table~\ref{tab:bestpar}, and the spectrum at
a few selected wavelength in Table~\ref{tab:modelfluxes}. The absorbed spectrum
of HZ~43 is represented with an accuracy of better than 0.5~\% over the full
43--180~\AA\ wavelength range by
\begin{eqnarray}
N_\lambda = \exp &(&-491.51/\lambda+12.277 \nonumber\\
                  &&-0.01418\lambda +11.8\times 10^{-6}\lambda^2 )
\label{eqn:hz43unconstrained1}
\end{eqnarray}
for model 1 and for model 2 with an accuracy better than 0.7~\% by
\begin{eqnarray}
N_\lambda = \exp &(&-509.12/\lambda+12.366 \nonumber\\
                  &&-0.01371\lambda +8.8\times 10^{-6}\lambda^2 ).
\label{eqn:hz43unconstrained2}
\end{eqnarray}

Note that (\ref{eqn:hz43unconstrained1})$-$(\ref{eqn:hz43unconstrained2}) 
should not be used outside this range.

\section{Discussion}

\subsection{Parameters of HZ~43}

\begin{figure}
\resizebox{\hsize}{!}{\includegraphics[angle=0]{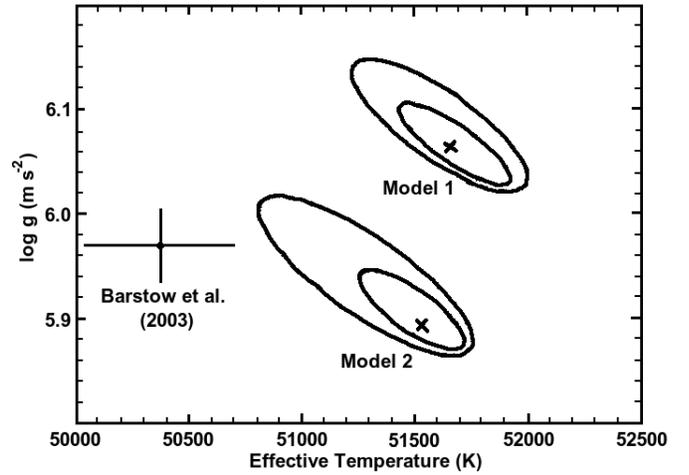}}
\caption{Contours of $\Delta\chi^2=2$ and $\Delta\chi^2=8$ in the effective
temperature - gravity plain of HZ~43A, for model 1 and
model 2. The cross indicates our best solution for each case. Also shown
are the nominal error bars on both parameters from the
analysis of \protect\citet{barstow2003}.}
\label{fig:ellips}
\end{figure}

Our lower limit to the hydrogen mass of $2.5$ to $3\times 10^{-12}$~\Msun\ is an
order of magnitude higher than the lower limit derived by \citet{barstow1995}
based on the EUVE continuum. We could derive this tighter limit because the
LETGS covers also shorter wavelengths, for which the continuum is very sensitive
to the hydrogen thickness (Fig.~\ref{fig:hlaag}). Our best-fit model is
indistinguishable from a pure hydrogen model, and even for our lower limit
hydrogen mass, above 50~\AA\ the differences with a pure hydrogen model are less
than a few percent.

One of the most recent alternative measurements of gravity and temperature of
HZ~43A were given by \citet{barstow2003} based on FUSE observations of the Lyman
series lines. They obtain values of $T_{\mathrm{eff}}=50\,380\pm 320$~K and
$\log g$ (m\,s$^{-2}$) $=5.97\pm 0.03$. These values are not consistent with our
model 1 or model 2 (Fig.~\ref{fig:ellips}), but they are closer to model 2. The
uncertainties given by \citet{barstow2003} correspond to the scatter between the
parameters derived from the individual fits of the only three FUSE spectra that
were  available, hence the nominal uncertainty may be quite uncertain by itself.
We also note that the differences between our best fit model and the model by
\citet{barstow2003} are less than 1~\% of the continuum level in the Lyman
series line cores, hence rather sensitive to uncertainties in scattered light
contributions or background subtraction.

Also, for model 1 the allowed range for the gravitational redshift is relatively
high, given that the best value is $30\pm 10$ to $30\pm 15$~km\,s$^{-1}$ (see
Sect.~\ref{gravred}). For the interstellar absorption column densities, our
model fits do not put strong constraints.

\subsection{Parameters of Sirius~B}

Until recently, the most accurate parameters of Sirius~B were given by
\citet{holberg1998}. These authors used the Lyman alpha line obtained by
IUE together with the EUVE spectrum to constrain the effective temperature and
surface gravity. With only the IUE observations the effective temperature is
known within $\pm 285$~K, but by including the EUVE spectrum this uncertainty
reduces to $\pm 100$~K.

In a recent paper, \citet{barstow2005} have decreased the formal error bars on
the effective temperature even further down to $\pm 37$~K by using
high-accuracy STIS spectra. However, the quoted uncertainty is only the
statistical uncertainty, and \citet{barstow2005} argue that the systematic
uncertainty on these numbers is hard to assess, mainly because there is little
else to compare with. A major reason of concern is the much larger effective
temperature (25\,193~K) found by \citet{barstow2005} as compared to the value
of 24\,790~K obtained by \citet{holberg1998}. While the surface gravity given
in both papers is almost equal and consistent within the error bars, this
temperature difference is $>4\sigma$ and causes the EUV flux at 300~\AA\ to
increase by a factor of 2.4 according to our own model calculations. Although
the absolute accuracy of the EUVE calibration has its limitations (see
Sect.~\ref{sect:euve_sir}), we believe that a factor of 2.4 cannot be easily
accommodated for. Moreover, \citet{barstow2005} also show that their best
normalisation of the G430L spectrum obtained with STIS has systematic
uncertainties larger than desirable. Given this problem with the EUV flux, we
prefer here the older \citet{holberg1998} parameters with the corresponding
error ranges. 

For model 1, the temperature is clearly higher (by 570~K) than the value found
by \citet{holberg1998}, while for model 2 it is consistent with the Holberg et
al. value within $2\sigma$ (only 180~K higher). For both models, $\log g$ is
consistent with \citet{holberg1998}.

According to \citet{holberg1998}, Sirius~B contains a small amount of helium
so pure hydrogen models are ruled out. For homogeneous H/He models, they found
$n_{\mathrm{He}}/n_{\mathrm{H}}=(4\pm 1)\times 10^{-6}$. Our upper limit to
the amount of helium of $<1.0\times 10^{-6}$ is well below that value. However,
the claim of detection of helium is based on the non-significant detection of
possible \ion{He}{ii} Ly$\delta$ and Ly$\epsilon$ lines in the EUVE spectrum,
as well as on the global fit to the EUVE spectrum. If we evaluate our models
for the parameters of \citet{holberg1998}, we find that there should be a deep
and sharp \ion{He}{ii} edge in the model near 230~\AA, with a depth of 13~\%.
The edge is very broad, and reaches half of its maximum depth at 100~\AA.
Clearly, such a deep edge is not observed in the EUVE spectrum, and the
systematic deviations from their best-fit model as shown in their Fig.~4 are
of the same order of magnitude. In fact, for wavelengths below the
\ion{He}{ii} edge the EUVE data show even a small systematic excess, pointing
to a lower helium abundance than adopted. We conclude that there is no
convincing evidence for a substantial amount of helium in Sirius~B.

As for HZ~43A, our models for Sirius~B do  not give additional useful constraints
for the interstellar absorption columns.

\subsection{The cut-off of the Lyman pseudo-continuum}

Based on our fits alone, it is not well possible to distinguish between model~1
(short Lyman pseudo-continuum cut-off) or model~2 (long cut-off), as both models
reproduce well the observed spectral ratio between Sirius~B and HZ~43A
(Fig.~\ref{fig:ratcomp}), albeit with different derived parameters for both
stars. When we look to those parameters (see previous subsection), it appears
that the derived temperature for Sirius~B and the surface gravity of HZ~43A are
in reasonable agreement with recent literature values only for model 2 (a long
cut-off). Model~2 therefore achieves a better consistency between analyses of
the UV/optical and soft X-ray ranges. Moreover, model 2 seems to match the
gravitational redshift of HZ~43A better.

The most direct test of the Lyman pseudo-continuum is provided by the far-UV
spectrum of Sirius B. The FUV spectrum was recorded with the far   ultraviolet
spectroscopic explorer (FUSE) on 2002 June 14, using medium resolution (MDRS)
and the SiC channel covering the range $\lambda\lambda$916-1100\,\AA. Model~2
with the long cut-off clearly provides the best match, though discrepancies up
to 10\% between the model and the FUSE spectrum  remain. In particular, the
observed spectrum reveals slightly broader high Lyman lines (Ly~$\gamma$ and
higher lines) than those predicted by model~2. The overall continuum flux level
is however well matched contrary to model~1. A detailed analysis of the FUSE
spectrum will be presented in a separate paper.

In summary, there seems to be more support for a long cut-off (model 2).

\subsection{A comparison with EUVE}

We have obtained fluxed, order-subtracted spectra of both stars from the public
EUVE archive\footnote{http://archive.stsci.edu/euve/search.php}. We sampled
these fluxed spectra on a grid with 5~\AA\ spacing using a spline fit, and
estimated the uncertainty on the flux point by looking to the r.m.s. variations
with respect to this fit in 5~\AA\ wide bins centred at the grid points (the
fluxed spectra from the public archive do not contain error estimates).  In
addition to these statistical uncertainties, we added systematic uncertainties
of about 2, 3.5 and 5 photons\,m$^{-2}$\,s$^{-1}$\,\AA$^{-1}$ for
$\lambda<150$~\AA, $150$~\AA\ $<\lambda<300$~\AA\ and $\lambda > 300$~\AA,
respectively. These are based upon a comparison of the spectra with smoothed
spectra on even larger scales of $\sim 50$~\AA. 

\begin{figure}
\resizebox{\hsize}{!}{\includegraphics[angle=-90]{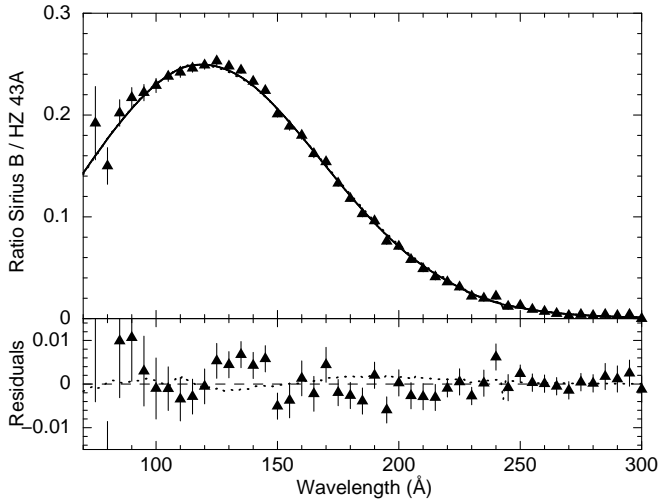}}
\caption{Ratio of the spectrum of Sirius~B to the spectrum of
HZ~43A. Data points with error bars: ratio as observed with EUVE. 
The solid line shows our best-fit model 2 (with a long wavelength
cut-off of the Lyman pseudo-continuum, and the dotted line our best-fit
model 1 (short cut-off), as discussed in Sect.~\ref{sect:results}.
Note that these fits are based solely upon the Chandra LETGS data,
not upon the EUVE data.
The lower panel shows the residuals of the observed ratio with
respect to model 2; the dotted line in that panel shows on the same scale
the differences between model 2 and model 1. Note that the first two data points
below 80~\AA\ are off-scale in this lower panel.}
\label{fig:ratio_euve}
\end{figure}

The ratio of these fluxed spectra of both stars is shown in
Fig.~\ref{fig:ratio_euve}. It is evident from this figure that the models that
we found using the observed Chandra LETGS ratios agree very well with the
observed EUVE data, even beyond the Chandra range for $\lambda>170$~\AA. Again,
these data cannot help to choose between model 1 and 2, although model 2
describes the data in the $170-250$~\AA\ range slightly better.

\begin{figure}
\resizebox{\hsize}{!}{\includegraphics[angle=-90]{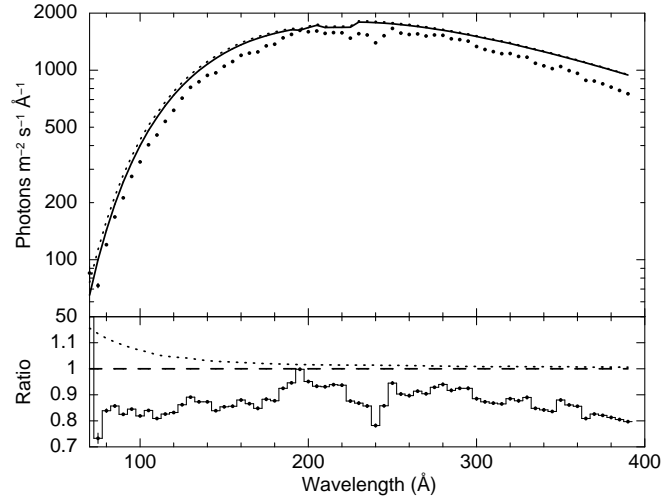}}
\caption{Fluxed EUVE spectrum of
HZ~43A. Data points: observed spectrum with EUVE. 
The solid line shows our best-fit model 2 (with a long wavelength
cut-off of the Lyman pseudo-continuum, and the dotted line our best-fit
model 1 (short cut-off), as discussed in Sect.~\ref{sect:results}.
The lower panel shows the ratio of the observed spectrum with
respect to model 2; the dotted line in that panel shows on the same scale
the differences between model 2 and model 1. }
\label{fig:hz_euve}
\end{figure}

In Fig.~\ref{fig:hz_euve} we compare the fluxed EUVE spectrum of HZ~43A with our
models 1 and 2. In the SW band (below 180~\AA) the EUVE flux is typically 15~\%
below our model flux, while in the MW band fluctuations up to 10~\% occur. Note
also the relatively large systematic fluctuations in both bands of up to a few
\% in the SW band to 5~\% in the LW band. 

\subsection{Comparison with \citet{beuermann2006}\label{sect:comparebeuermann}}

We have compared our model calculations with \citet{beuermann2006} by evaluating
our model using exactly the same parameters as obtained by these authors (their
Table~2). We show this comparison in Table~\ref{tab:comparebeuermann}. We have
used here the long cut-off of the Lyman pseudo-continuum (see
Sect.~\ref{sect:pseudo}).

It should be noted that Table~2 and Table~3 of \citet{beuermann2006} contain
errors, as explained in an erratum on that paper (in press). This erratum was
triggered by our present results. Accordingly, we used their updated
temperatures of 24897 and 51111~K for Sirius~B and HZ~43A, as well as the
updated fluxes (Table~1 of the erratum).

\begin{table}
\begin{center}
\caption{Comparison of model spectra for the same set of parameters (those of
\citealt{beuermann2006}). Fluxes are in photons\,m$^{-2}$\,s$^{-1}$\,\AA$^{-1}$
and include interstellar absorption.}
\begin{tabular}[t]{rcccccc}
 \hline
 \hline
 & \multicolumn{3}{c}{HZ~43A} & \multicolumn{3}{c}{Sirius~B} \\
$\lambda$ (\AA) & TMAP & Tlusty & ratio  & TMAP & Tlusty & ratio \\ 
          48 &    3.00&    2.89 & 0.963 &  -    & 0.20  &  -    \\
          60 &   21.3 &   20.95 & 0.984 & 2.45  & 2.23  & 0.910 \\
          70 &   63.2 &   62.58 & 0.990 & 9.38  & 8.75  & 0.933 \\
          80 &  139.  &  137.4  & 0.988 & 25.2  & 23.7  & 0.940 \\
          90 &  249.  &  246.7  & 0.991 & 53.0  & 49.8  & 0.940 \\
         100 &  388.  &  384.5  & 0.991 & 92.0  & 86.3  & 0.938 \\
         125 &  795.  &  792.5  & 0.997 & 201.  & 189.8 & 0.944 \\
         160 & 1310.  & 1306.   & 0.997 & 237.  & 220.6 & 0.931 \\
        1300 & 2652.  & 2630.   & 0.992 & 82000 & 78660 & 0.959 \\
        4600 &  -     & 111.4   &  -    & 6611  & 6450  & 0.976 \\
        5450 & 70.18  & 69.11   & 0.985 &  -    & 4192  &  -    \\
\hline       
\end{tabular} 
\label{tab:comparebeuermann}
\end{center}
\end{table}

For HZ~43A there is an excellent agreement between both codes; only at the
shortest wavelength listed (48~\AA), there is a small 4~\% difference.
However, for Sirius~B there are large differences. It is striking that at all
wavelengths our predicted flux is smaller than the flux given by
\citet{beuermann2006}, also because we used exactly the same interstellar
absorption column as well as normalisation $R^2/d^2=4.877\times 10^{-21}$ as
these authors. We verified that the (unabsorbed) and integrated spectrum of
our model obeys with high precision the normalisation condition that $\int
F(\nu) {\mathrm d}\nu = \sigma T_{\mathrm {eff}}^4$ with $\sigma$ the
Stefan-Boltzmann constant and $F(\nu)$ the emitted surface flux. As we both
use the same effective temperature of 24\,897~K, the conclusion must be that
the spectrum for Sirius~B as calculated by \citet{beuermann2006} is probably
not correct. 

Independently, we compare the ratio of the spectra of both stars as calculated
by \citet{beuermann2006} to the ratio that we measured with the LETGS. Again,
the measured ratio is smaller by on average a factor of $0.941\pm 0.008$. The
reason for this discrepancy is not clear, but we note that our ratio for the
LETGS spectra is consistent with the ratio obtained from the EUVE spectra. 

Remarkably, our own normalisation constant for Sirius~B for the same value of
$g$ and $T_{\mathrm{eff}}$ is 7.3~\% higher than the value given by
\citet{beuermann2006}, the main difference being that we use a value of $R$
based on the measured gravitational redshift instead of the spectral modelling
derived effective gravity. If we would have used our own normalisation in
Table~\ref{tab:comparebeuermann}, the discrepancy would have been smaller.

\subsection{Effective area comparison}

\begin{figure}
\resizebox{\hsize}{!}{\includegraphics[angle=-90]{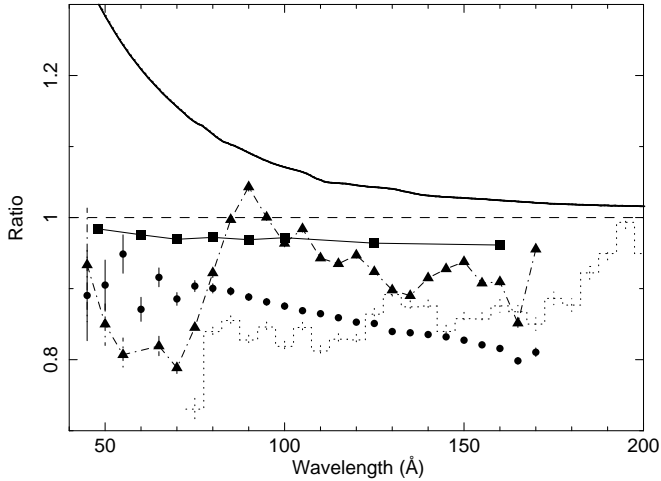}}
\caption{Fluxes of HZ~43A with respect to the model flux of our model 2
(long cut-off). Solid line: model 1; circles: results of a spline fit
to the LETGS spectrum using the old SRON effective area calibration
based on work in 2000; dashed histogram: EUVE flux; squares:
\citet{beuermann2006}; dash-dotted line, triangles: results of a spline
fit to the LETGS spectrum using the standard CXC CIAO pipeline.}
\label{fig:relarea}
\end{figure}

As the effective area of the LETGS depends on details such as pha (pulseheight)
selection, the spectral order, or the width of the spectral extraction box, it
is not very usefull to give the effective area here. Instead, we compare here
directly  model spectra. This comparison is shown in Fig.~\ref{fig:relarea}.

A comparison between model 1 and 2 shows that in particular at the shorter
wavelengths the differences are large: at 40~\AA, model 1 predicts 38~\% more
flux than model 1. This difference is solely due to the adopted value of the
Lyman pseudo-continuum cut-off that affects the model calculations for Sirius~B.
As we fit both stars together, however, the different spectrum for Sirius~B
implies then a different solution for HZ~43A, as the ratio of both spectra is
constrained by the observed ratio with Chandra. The difference between
both models then must be found through a comparison with other data, for
instance temperatures and effective gravities determined from detailed
line fitting.

The differences between the present model 2 and our older (2000) effective
area based on work by J. Heise and J. Kaastra are less than 20~\% for
$\lambda>80$~\AA, and differ by 10~\% below 80~\AA. Compared to
Model 1, there is a constant offset of about 20~\% for $\lambda>80$~\AA;
for the shorter wavelengths, the effective area in the 2000 version was not
based upon HZ~43 but on matching blazar spectra from shorter wavelengths, so it
is not surprising that the differences are larger for shorter wavelengths. 

The model flux given by \citet{beuermann2006} is on average 3~\% lower than the
flux that we derive for model 2, although the shape agrees within 1~\%.

We also compare our model to the model flux derived directly from the observed
LETGS spectrum of HZ~43A as processed and modelled using the official CXC
software (instead of our local SRON software). We used CIAO version 4.0 with
CALDB version 3.4.2. The spectrum was fitted using a spline, similar to what we
described in Sect.~\ref{sect:letgs}, and we show in Fig.~\ref{fig:relarea} the
ratio of this fluxed spectrum to our model 2 with the dash-dotted line. Above
90~\AA, there is good agreement in shape with model 1 (apart from a 10~\% flux
difference), but for shorter wavelengths there is a strong dip: between
$70-90$~\AA, there is a relative change of almost 25~\% between our models and
the CXC-based model. The large scale (tens of \AA) fluctuations with an
amplitude of a few percent as compared for instance to our old (2000)
calibration and also to \citet{beuermann2006} are not very surprising, as we
fudged effectively the effective area to get the observed spectrum agree by
definition to the predicted model.

Finally, the dotted line shows the comparison of the EUVE fluxed spectrum
discussed earlier to our model 2. Over the wavelength band of the LETGS, the
difference with model 2 is $\sim 15$~\%.

\subsection{Uncertainty of the model flux}

\begin{figure}
\resizebox{\hsize}{!}{\includegraphics[angle=-90]{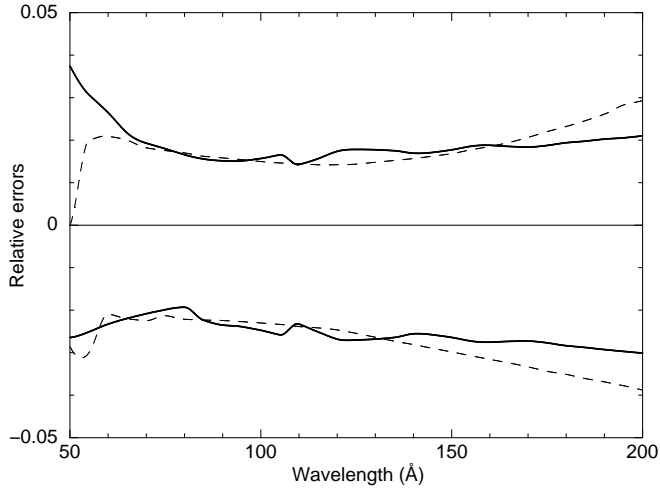}}
\caption{Relative uncertainty on our model spectra for Sirius~B (dashed lines)
and HZ~43A (solid lines). All acceptable solutions with $\Delta\chi^2 <
\chi^2_{\min}+2$ are bounded by these lines (which we scaled to our
best solution for each star). Results shown are for model 2; the results for
model 1 are quite similar.}
\label{fig:accu}
\end{figure}

The statistical uncertainty of our model spectrum for both stars is shown in
Fig.~\ref{fig:accu}. Although not as good as the flux limits in the optical
band (1~\% accuracy), we still can reach an accuracy of better than 5~\% on
the absolute flux of both stars. As we will show in another paper
\citep{paper3}, we can even reduce this uncertainty to 3--4~\% by using
additional constraints at high energies. Here we will carefully asses the
possible size of systematic uncertainties in our model spectra.

We follow the same approach as \citet{paper1}. We estimate the known
systematic uncertainties and present them in Table~\ref{tab:syserr}. For each
relevant factor, we  estimate its magnitude for a range of characteristic
wavelengths, covering the LETGS range between 50--170~\AA. These numbers are
given in the table. Then we put each error in one of two specific categories. 

For category "c", the systematic uncertainties at different wavelengths are more
or less correlated. Example: an error in the interstellar absorption column
density will lead to correlated deviations. As a rough approximation, such
deviations can be approximated by a power law in wavelength. We can assess for
these "correlated" errors how they will affect the normalisation and slope of
this power law written in the form $f(\lambda) = A(\lambda/100)^\alpha$ by
"fitting" the estimated systematic uncertainties (all having the same sign)
directly to such a power law shape. 

For category "u", the systematic uncertainties at different wavelengths are
uncorrelated over the full LETGS wavelength scale. A good example are the
statistical errors on the Sirius~B to HZ~43A ratio. In those cases, we
estimated $\Delta A$ from the typical slope and normalisation that we would
obtain if the systematic uncertainties at different wavelengths had random
signs. In general, for this type of uncertainty the direct effect on $A$ and
$\alpha$ is smaller than for correlated errors (category "c").

More details on the individual systematic uncertainties are given in
Appendix~A. In summary, the most important factor is the uncertainty in the
spectral modelling. This can give a flux difference at 100~\AA\ of 1~\% and a
difference in slope of 0.022. However, these values are well within the
uncertainty range that we found for the spectrum of HZ~43
(Fig.~\ref{fig:accu}).

\begin{table*}
\begin{center}
\caption{Summary of systematic and statistical uncertainties in the Sirius~B and
HZ~43A spectra. }
 \begin{tabular}[t]{r|ccc|ccc|cccc|c}
 \hline
 \hline
 $\lambda$      &  a   &  b     &  c    &   d   &   e   &  f    &   g   &   h   &  i    &   j   &  k    \\
\hline
50              & 0.000 & 0.000 & 0.000 & 0.005 & 0.027 & 0.007 & 0.072 & 0.000 & 0.000 & 0.005 & 0.192 \\
60              & 0.001 & 0.000 & 0.001 & 0.002 & 0.014 & 0.007 & 0.011 & 0.002 & 0.000 & 0.005 & 0.122 \\
80              & 0.003 & 0.001 & 0.001 & 0.001 & 0.005 & 0.007 & 0.004 & 0.002 & 0.000 & 0.005 & 0.020 \\
100             & 0.005 & 0.001 & 0.001 & 0.000 & 0.003 & 0.007 & 0.002 & 0.004 & 0.000 & 0.005 & 0.018 \\
120             & 0.007 & 0.002 & 0.001 & 0.000 & 0.005 & 0.007 & 0.002 & 0.005 & 0.001 & 0.005 & 0.012 \\
140             & 0.010 & 0.004 & 0.002 & 0.000 & 0.006 & 0.007 & 0.002 & 0.005 & 0.002 & 0.005 & 0.010 \\
160             & 0.011 & 0.006 & 0.002 & 0.000 & 0.008 & 0.007 & 0.003 & 0.006 & 0.006 & 0.005 & 0.012 \\
170             & 0.012 & 0.007 & 0.002 & 0.000 & 0.010 & 0.007 & 0.051 & 0.008 & 0.010 & 0.005 & 0.031 \\
\hline
\multicolumn{1}{c|}{type}&
                    c   &   c   &   c   &   c   &   c   &   u   &   c   &   c   &   u   &   c   &   u   \\
\multicolumn{1}{c|}
{$\Delta\alpha$}& 0.010 & 0.006 & 0.001 & 0.003 & 0.022 & 0.006 & 0.008 & 0.005 & 0.007 & 0.000 & 0.027 \\
\multicolumn{1}{c|}
{$\Delta A$}    & 0.006 & 0.003 & 0.001 & 0.001 & 0.004 & 0.002 & 0.004 & 0.004 & 0.002 & 0.005 & 0.006 \\
\end{tabular} 
\label{tab:syserr}
\end{center}
{\sl Interstellar absorption:} \\
$^a$ Uncertainty in the absorption cross sections \\
$^b$ Uncertainty in the ionisation correction \\
$^c$ Neglect of metals in the opacity \\
{\sl Atmosphere modelling:} \\
$^d$ Neglect of Compton scattering \\
$^e$ Uncertainties in atmosphere models \\
$^f$ Interpolation errors \\
{\sl Spectral extraction:} \\
$^g$ Background uncertainty \\
$^h$ Background contamination by source cross-dispersion distribution \\
$^i$ Higher spectral order subtraction \\
$^j$ Off-axis vignetting Sirius~B \\
$^k$ Statistical uncertainty spectra \\
\end{table*}

\section{Conclusions}

We have modelled carefully the X-ray spectra of Sirius~B and HZ~43A using the
most sophisticated spectral models presently available. By using the observed
spectral ratio of both stars with LETGS as constraints, together with some other
constraints, we have obtained models for each star that predict the flux in the
$50-170$~\AA\ band to better than 5~\%. This allows also to calibrate the
relative and absolute effective area of the LETGS in the same wavelength band
with the same accuracy. 

The model spectrum for Sirius~B appears to depend significantly upon the
assumption made about the cut-off of the Lyman pseudo-continuum. This affects
our solutions in two ways. First, the best fit parameters for both stars depend
upon this choice, and secondly the absolute flux of both stars at the shorter
wavelengths depends strongly upon it. Accordingly, we have investigated here two
classes of models for Sirius~B: model 1 with a short cut-off of the Lyman
pseudo-continuum, model 2 with a long cut-off. From our X-ray data alone we
cannot distinguish between both models. However, it appears that only for model
2 (a long cut-off) the derived temperature and far UV flux of Sirius~B, and the
temperature, surface gravity and gravitational redshift of HZ~43A agree much
better with recent analyses of the UV and optical spectra.

For both stars, we can put tight upper limits to the amount of helium or a tight
lower limit to the thickness of a hydrogen layer. In fact, both stars show
to agree remarkably well with a pure hydrogen atmosphere. The values that we
derive for the surface gravity and temperature of both stars have an accuracy
comparable to what has been achieved by detailed line profile fitting, apart
from the larger differences between  model 1 and model 2.

\begin{acknowledgements}

We thank Jay Holberg for helpfull comments on this paper, Vadim Burwitz for
discussions about the Chandra data analysis, and Thomas Rauch for discussions on
the difference between the TMAP and Tlusty calculations. This research has made
use of data obtained from the Chandra Data Archive and software provided by the
Chandra X-ray Center (CXC) in the application package CIAO. SRON is supported
financially by NWO, the Netherlands Organization for Scientific Research. 

\end{acknowledgements} 
 
\bibliographystyle{aa}
\bibliography{wd}

\begin{appendix}

\section{Breakdown of systematic uncertainties}

\subsection{Interstellar absorption}

We have used the interstellar absorption cross sections of \citet{rumph1994}.
Replacing the atomic cross sections with those of \citet{verner1995} gives small
differences, as listed in column~(a) of Table~\ref{tab:syserr}. 

Also, we use slightly different ionisation fractions of hydrogen and helium as
compared to \citet{beuermann2006}. Replacing our ionisation fractions by those
of \citet{beuermann2006} yields the differences in column~(b).

Further, in the absorption model of \citet{rumph1994} all elements other than
hydrogen and helium are ignored. We estimate this effect by comparing the
transmission with and without metals using the {\sl hot} absorption model of
SPEX with the appropriate parameters. The differences are very small and shown
in column~(c).

\subsection{Atmosphere models}

The effects of ignoring Compton scattering has been estimated from the study
of \citet{suleimanov2006} and are listed in column~(d). 

Using the adopted model atmosphere for HZ~43A, we estimated the internal
accuracy of model atmosphere calculations by allowing for different treatment  
of the hydrogen opacity (line broadening, extension of Lyman pseudo-continuum)
and using different discretisation in frequency and depth. The cumulative
systematic uncertainties are listed in column~(e). These estimates are similar
to the differences found in calculating a model with Tlusty and TMAP using the
same parameters (see for instance Table~\ref{tab:comparebeuermann}, the column
for HZ~43A).

In our fitting procedure, we calculate atmosphere models by interpolating
models on a three-dimensional grid (temperature, gravity and helium abundance
or hydrogen column mass). We estimate the uncertainties corresponding to this
interpolation by comparing a few interpolated models with exact calculations.
We list these in column~(f). 

\subsection{Spectral extraction\label{sect:extraction}}

The background towards both stars was determined from boxes below and above the
spectral extraction region. The background of the HRC-S detector is not
completely flat. For typical HRC-S extraction regions, the background below and
above the spectrum may differ by 2--3~\%. As a conservative estimate we
therefore can assume that the average background uncertainty in the source
extraction region is less than 1~\%. As Sirius~B is the faintest of both stars,
this uncertainty is the strongest for that source. We list the corresponding
uncertainty in column~(g).

HZ~43A is a very strong source. Inspection and modelling of the empirical
distribution of photons in the cross-dispersion direction shows that at the
longest wavelengths the background can be enhanced by 10--15~\% due to scattered
photons of HZ~43A. We have not taken this into account in our background
subtraction, but the effect can be easily estimated, see column~(h).

The higher spectral order contribution to the HZ~43A spectrum is small, less
than 3~\% for all wavelengths. Adopting conservatively a maximum uncertainty
of 30~\% on the estimated higher order calibration relative to the first
order, we determined the maximum uncertainty due to higher order subtraction
as listed in column~(i).

Our observation of Sirius~B was taken slightly off-axis with respect to HZ~43A
by about 1\arcmin. According to the Proposer's Observatory Guide of Chandra,
the main effect of this off-axis angle is due to vignetting of the Chandra
mirror. At the low energies that are relevant for white dwarfs and this
off-axis angle, the vignetting effect is less than 0.5~\% and constant as a 
function of wavelength, see column~(j).

Finally, we list the statistical errors on the observed ratio of both stars
in column~(k).

\end{appendix}

\end{document}